\documentclass[fleqn,usenatbib]{mnras}

\usepackage{newtxtext,newtxmath}

\usepackage[T1]{fontenc}

\usepackage{setspace}
\usepackage{multirow}
\usepackage{hyperref}

\DeclareRobustCommand{\VAN}[3]{#2}
\let\VANthebibliography\thebibliography
\def\thebibliography{\DeclareRobustCommand{\VAN}[3]{##3}\VANthebibliography}

\usepackage{graphicx}	
\usepackage{amsmath}	

\title[WHTZ~1: A PN not a cocoon from HD~185806]{WHTZ~1: A high excitation Planetary Nebula not a gaseous cocoon from runaway star HD~185806}

\author[Q.A. Parker et al.]{Quentin A. Parker,$^{1,2}$\thanks{E-mail: quentinp@hku.hk (QAP)}
Pascal Le D\^u,$^{3,4}$
Andreas Ritter,$^{1,2}$ 
Peter Goodhew,$^{5}$
Sakib Rasool,$^{6}$
\newauthor
Stephane Charbonnel,$^{3}$
Olivier Garde,$^{3}$
Lionel Mulato,$^{3}$
and Thomas Petit$^{3}$
\\
$^{1}$Department of Physics, CYM Physics Building, The University of Hong Kong, Pokfulam, Hong Kong PRC \\
$^{2}$The Laboratory for Space Research, The University of Hong Kong, Cyberport 4, Hong Kong PRC\\
$^{3}$2SPOT, 38690 Chabons, France\\
$^{4}$Kermerrien Observatory, 29840 Porspoder, France\\
$^{5}$W4 3EQ, London, UK\\
$^{6}$Manchester, UK\\
}

\date{Accepted XXX. Received YYY; in original form ZZZ}

\pubyear{2022}

\begin{document}
\label{firstpage}
\pagerange{\pageref{firstpage}--\pageref{lastpage}}
\maketitle

\begin{abstract}
We present evidence that the nebular cocoon and bow-shock 
emission nebula putatively and recently reported as deriving from the 
9$^{th}$ magnitude "runaway" star HD~185806 is the previously 
discovered but obscure planetary nebula WHTZ~1 (Ra~7). It has a Gaia DR3 G$\sim$16 blue 
ionizing star at its geometric centre. We present imagery, 
spectroscopy, other data and arguments to support that this emission source 
is a high excitation Planetary Nebula not a stellar wind bow shock.
\end{abstract}




\section{Introduction}

Miss-identification, false discoveries and re-assignments of celestial 
objects from one class to another are common as new knowledge and data of
higher resolution, sensitivity and wavelength coverage become available. 
The field of Planetary Nebula (PN) research, history is 
littered with many examples of objects that have
been re-discovered and re-identified as PNe only to later be 
re-classified as one of the many PN mimics that exist (such as HII regions, Wolf-Rayet 
shells, Herbig-Haro nebulae, YSO's, reflection nebulae, supernova 
remnants, nova shells etc). See 
\citet{2010PASA...27..129F} and the recent
review by \citet{Parker2022} for further details. 
Refer also to \citet{2021ApJ...918L..33R} for a PN candidate, Pa~30, re-classified 
recently as a supernova remnant of the Chinese "guest star" of 1181~AD. 

It is much rarer to find examples where a previously reported PN candidate 
is later ascribed to another source type but then needs to be moved back to its 
original classification. 

One recent case concerns the discovery ({\it{sic}}) of a nebula cocoon reported
as deriving from runaway star HD~185806 \citep{2022MNRAS.515.1544S}.
The star is a long period variable also designated as V~1279 Aql, where the variability
is due to pulsation not binarity \citep[e.g.][]{1966ARA&A...4...19S}. We present 
on-record historical context that shows that the adjacent uncovered gaseous emission 
region is the previously discovered source WHTZ~1 and observational evidence to re-assign this source back to its original PN status.

\subsection{Historical context}

The nebulosity was first noticed as a faint emission region on photographic 
B-band imagery from the POSS-I survey by \citet{1999ASPC..168..142W} and given 
the name WHTZ~1. It 
was found as a serendipitous by-product when looking for hidden galaxies 
behind the Milky way. They identified it as a PN candidate with a 
morphology "typical" of many PNe and with a blue central star (CSPN). The 
source was never incorporated into the SIMBAD astronomical 
database \citep{2000A&AS..143....9W} at CDS, Strasbourg.

The object was independently "re-found" by French PN amateur 
astronomer Thierry Raffaelli in 2014 and called "Ra~7" where it was added
to the dedicated on-line database created to record and present their 
work\footnote{ 
\url{http://planetarynebulae.net/EN/page\_np.php?id=237}} and see Le D\^u et al. (in
press). The object appears as "Ra~7" in SIMBAD. It was also re-discovered by the Deep Sky Hunters team led by Matthias 
Kronberger and Dana Patchick \citep[e.g.][]{2016JPhCS.728g2012K},
based in part on assessments of its WISE MIR imagery \citep{2010AJ....140.1868W}. It 
was ingested into the "HASH" database\footnote{\url{http://hashpn.space/}} as a likely PN in 2015, with unique HASH ID~4418. HASH is the 
current "gold standard" consolidated  on-line catalogue of all known PN in our Galaxy and Magellanic clouds \citep{2016JPhCS.728c2008P} having evolved from the previous "MASH" surveys, \citet{2006MNRAS.373...79P}, \citet{2008MNRAS.384..525M}. 

It has also been reported in several editions of the French popular  
"L'Astronomie" magazine in February 2015, vol. 129, No. 80, 
p.42\footnote{\url{https://ui.adsabs.harvard.edu/abs/2015LAstr.129b..42A/}};  
February 2016, vol. 130, No. 91, 
p.26\footnote{\url{https://ui.adsabs.harvard.edu/abs/2016LAstr.130b..26A/}} and finally
\citet{2017LAstr.131b..46L}. 


The source has now re-appeared in the literature as a bow-shock cocoon nebula of star HD~185806 in \citet{2022MNRAS.515.1544S}.

\section{Methods and Observations}

We provide accumulated multi-wavelength imagery, new optical 
spectroscopy, identification of the actual ionising star, 
associated data, salient characteristics and other evidence to confirm the source 
as a "True"  high excitation PN following the precepts outlined in 
\citet{2010PASA...27..129F} and \citet{Parker2022}. The star HD~185806 
is merely a chance, closely projected neighbour that is unrelated to the 
actual emission observed.

\subsection{More recent deep amateur narrow-band imagery}

\begin{figure*}
	\includegraphics[width=18cm]{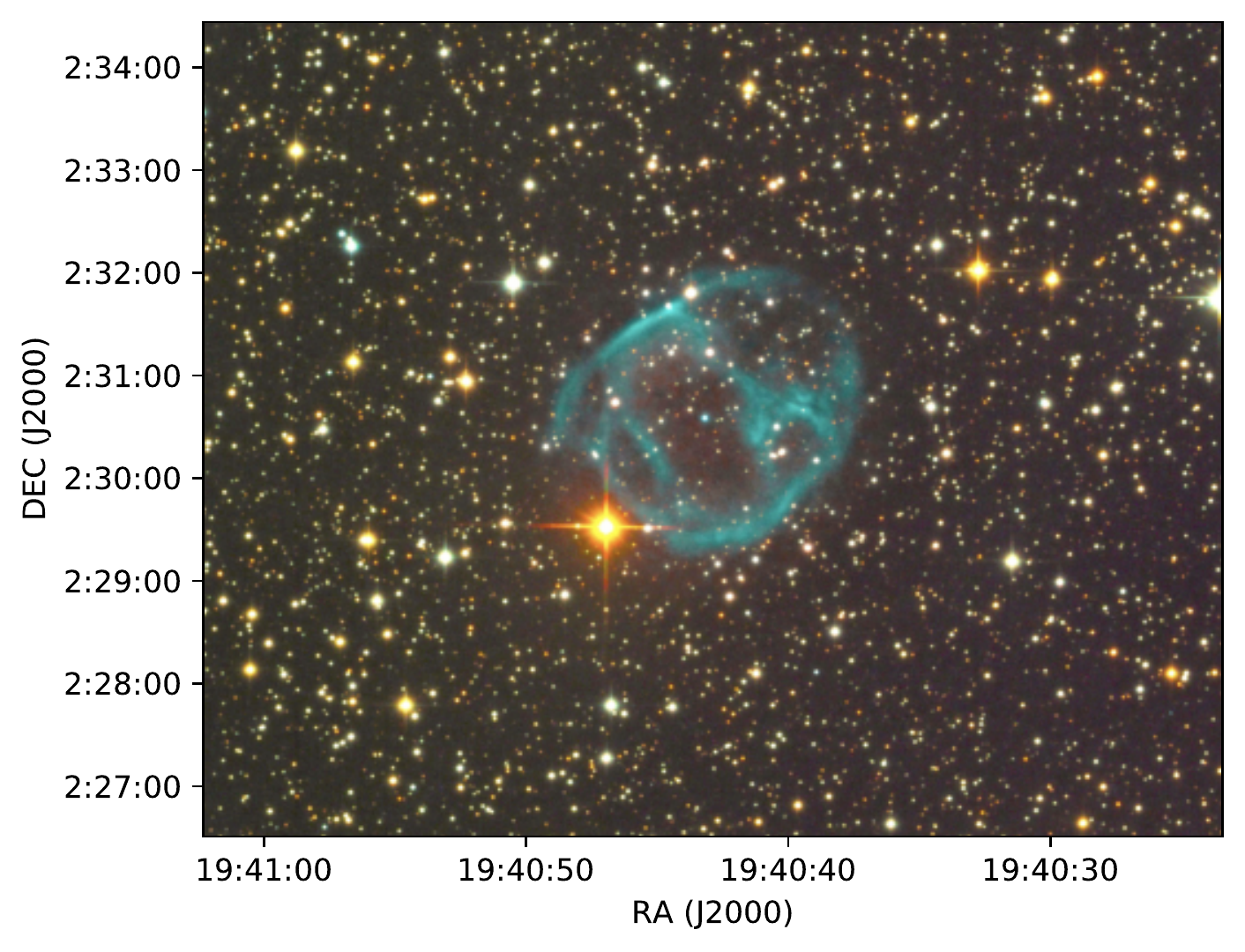}
    \caption{Combined 10$\times$8~arcmin narrow-band (30\AA) image of WHTZ~1 
    comprising combined H$\alpha$ images of 280~minutes (red-channel), combined 
    [OIII] images of 360~minutes (green channel) and then 3 broad-band 
    Astrodon Gen2 RGB filter images for 100~min each respectively (blue 
    channel). This gave a total on-source exposure time of 15.7 hours. In
    the figure North is up and East to the left. The typical seeing was 
    reported as 0.8-1.2~arcseconds. The blue CSPN is clear at 
    the nebula's geometric centre.}
    \label{fig:chart32}
\end{figure*}

Deep, narrow-band, optical imagery of the candidate PN was taken with the 
Chart32 "Chilean advanced robotic 32inch telescope" in 2017. The 
resultant 10$\times$8~arcmin RGB composite narrow-band image is shown in 
Fig~\ref{fig:chart32}, taken 
directly from the Chart32 open-access website: \url{www.chart32.de}. This 
colour image can be directly compared to Figs. 1,2 \& 3 in 
\citet{2022MNRAS.515.1544S}. 

The Chart32 RGB colour image was formed from co-adding narrow-band H$\alpha$ images (bandpass 30\AA) of 280~minutes total duration (red 
channel), [OIII] images of 360~minutes total duration (green channel) 
and then 3 broad-band Astrodon Gen2 RGB filter images (blue channel) for 
a total of 100~min each respectively. This gave a total on-source 
exposure time of 15.7 hours.  The typical seeing was reported as 0.8-1.2~arcseconds and 
the data were acquired with an 81cm f/7 Astrooptik Keller Cassegrain 
telescope with the images processed by Johannes Schedler of the Chart32 
team. 

\begin{figure}
	\includegraphics[width=8.5cm]{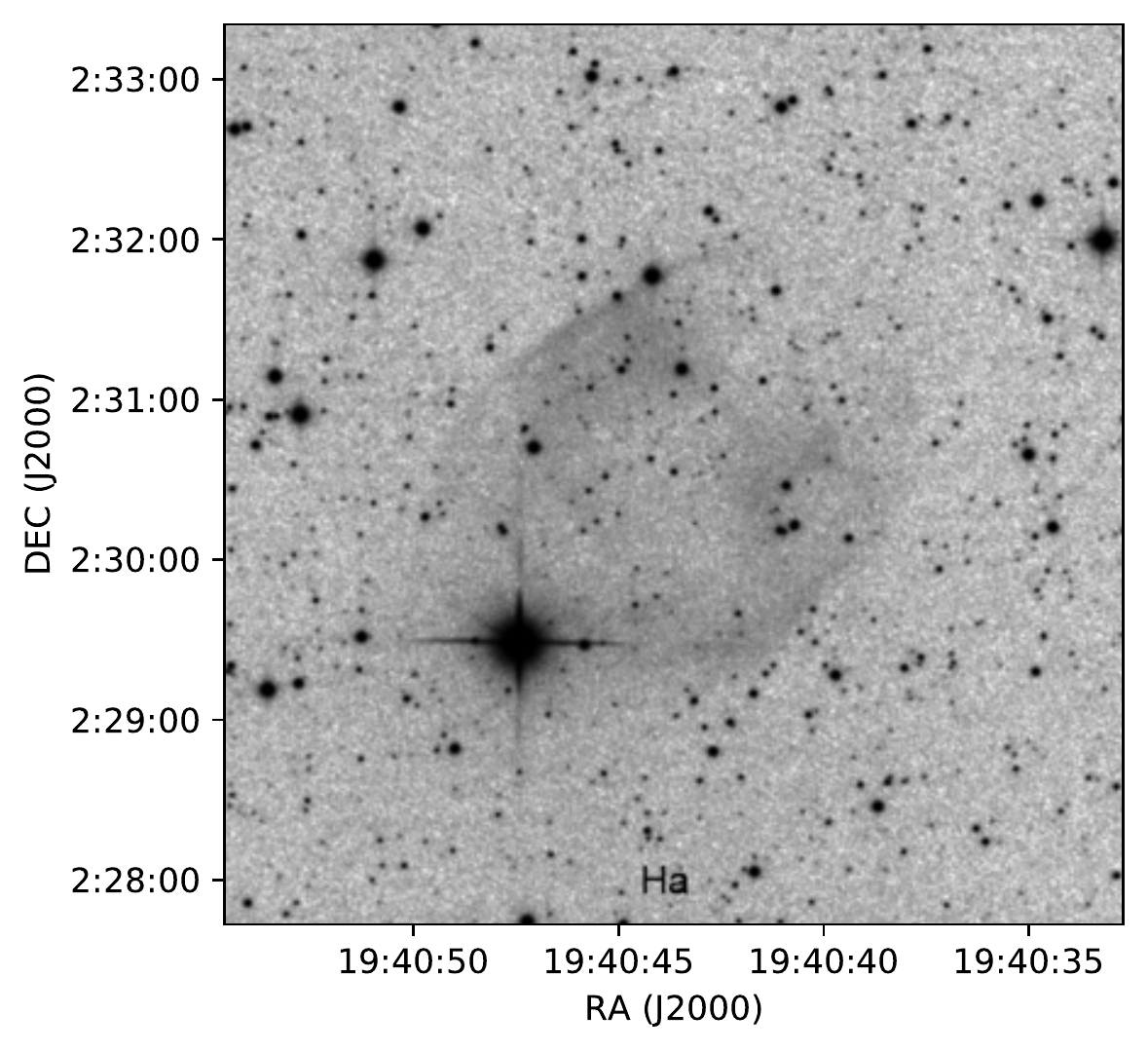}
    \caption{Chart32 H$\alpha$ 5$\times$5~arcmin image for WHTZ~1 showing that the H$\alpha$ closely follows that of the [OIII] but is fainter and less distinct in character for the equivalent exposure time. North is up and East to the left.}
    \label{fig:WHTZ1-Halpha}
\end{figure}

In Fig~\ref{fig:WHTZ1-Halpha} we also show the 5$\times$5~arcmin Chart32 H$\alpha$ 
image to show H$\alpha$ emission is present. The overall form of the 
stronger signal seen in [OIII] is there but is somewhat less distinct in character.

The main body of the WHTZ~1 nebula is oval in shape with a major axis $\sim$193~arcseconds 
across and a minor axis of $\sim$134~arcseconds. The opposing  edges across the 
minor axes have enhanced intensity and there are internal 
striations approximately perpendicular to these edges. There is a 
CSPN located almost exactly at the geometric centre of the 
nebula, like many typical PNe of elliptical morphology.

\begin{figure}
	\begin{center}
	\includegraphics[width=8.5cm]{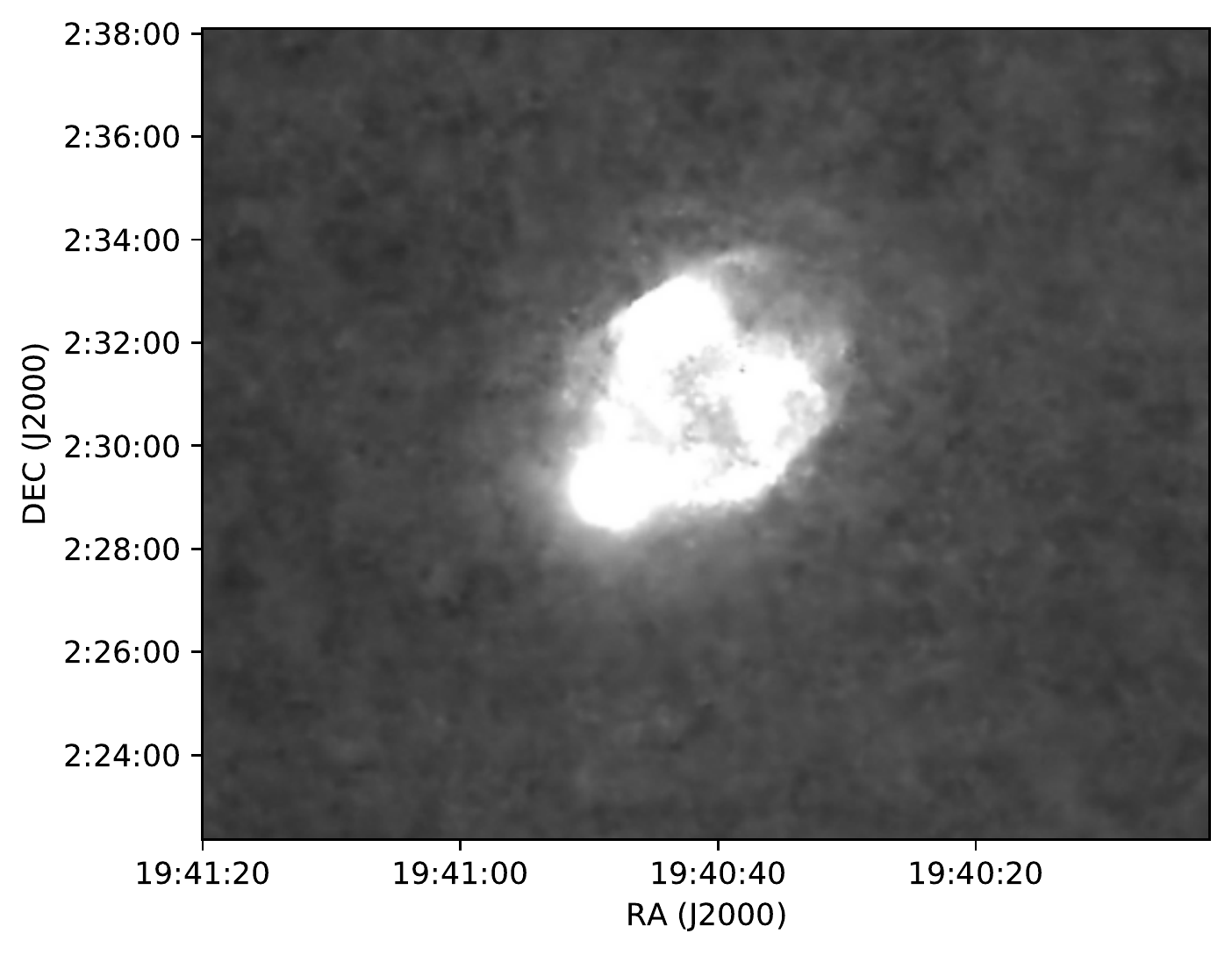}
    \caption{Large area, contrast enhanced, deep H$\alpha$ 20$\times$16~arcmin image for 
    WHTZ~1, with stars removed for clarity. Clear evidence of a faint $\sim$10$\times$7~arcmin
    H$\alpha$ halo is seen surrounding the main nebula. North is up and East to the left. The 
    image derives from 40~hours and 25~mins of combined narrow-band exposures.}
    \label{fig:Deep-Halpha}
    	\end{center}
\end{figure}

We have supplemented these data with very deep narrow-band H$\alpha$ 
imaging\footnote{see https://www.imagingdeepspace.com/observatories.html)}. A combined total of 40~hours and 25~mins H$\alpha$ imaging were combined from observations on 
August 17$^{th}$ and September 8,9,10,13,15,16$^{th}$ 2022 using two 
separate automated telescopes in the UK (Celestron C14 EdgeHD telescope
with ZWO ADI6200MM Pro camera and a Chroma H$\alpha$ 3~nm bandpass filter for 13~hours and 10~mins observations)
and Spain (Twin APM LZOS 152/1200 telescopes with QSI6120 cameras and an Astrodon H$\alpha$ 3~nm bandpass filter for 27~hours and 15mins of observations). In Fig~\ref{fig:Deep-Halpha} we present the resultant contrast enhanced 
20$\times$16~arcmin H$\alpha$ image of WHTZ1 (stars removed for clarity) obtained over these $\sim$40~hours of combined imagery. A faint outer halo can be discerned which is $\sim$10~arcmins across the major axis. Such an extensive, faint halo is difficult to reconcile with any bow-shock scenario.

Deep amateur narrow-band 
exposures of PN and PN candidates are now providing highly competitive imagery, 
matching, and in many cases exceeding, the best professional efforts previously 
available \citep{Parker2022}.

\subsection{Mid Infrared Imagery}

\begin{figure}
	\begin{center}
	\includegraphics[width=8.5cm]{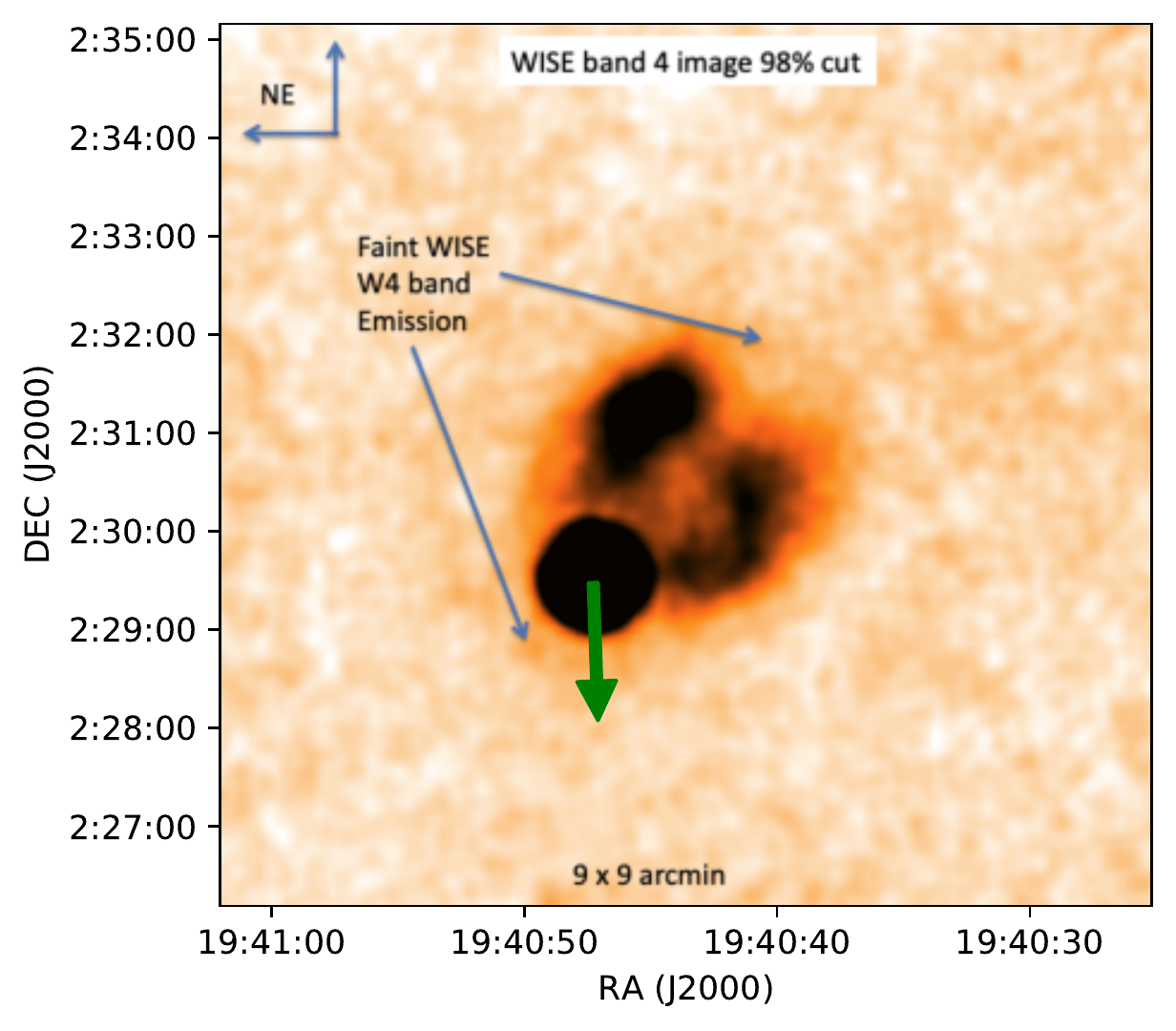}
    \caption{A 9$\times$9~arcmin WISE W4 22$\mu$m image for WHTZ~1 and 
    HD~185806. It shows the main opposing arc like structures and hollow 
    interior. Annotation indicates the faint MIR emission of the outer 
    oval nebula to the SE and NW reflecting what is also seen in the 
    optical. The green arrow marks the almost directly Southern direction of the Gaia proper 
    motion vector for HD~185806.}
    \label{fig:WISE}
    	\end{center}
\end{figure}

In Fig.~\ref{fig:WISE} we show a 9$\times$9~arcmin WISE W4 22$\mu$m \citep{2010AJ....140.1868W}
image for WHTZ~1 and HD~185806. The well defined emission region is prominent and has two thick, 
opposing arc-like structures that
follow the enhanced [OIII] edges seen in the optical. The interior is of much 
lower intensity suggesting a cavity. The Northern MIR emission arc is more 
prominent. The nebula can also be seen in the W3 12$\mu$m image, contrary to 
what was reported by \citet{2022MNRAS.515.1544S}. Annotation indicates the 
faint MIR emission of the outer oval nebula to the SE and NW reflecting 
what is also seen in the optical.
This compact 3$\times$2~arcmin MIR emission region is isolated. 
The proper motion vector for HD~185806 (see later) is  added,  taken from 
Gaia DR3 data for this star, e.g. \citet{2016A&A...595A...1G}. The prominence of 
HD~185806 in the MIR is also clear.

\subsection {Optical Spectroscopy}
No low resolution spectroscopy of the nebula was 
 presented by \citet{2022MNRAS.515.1544S} who obtained high resolution 
 echelle spectroscopy that covered the H$\alpha$ + [NII] lines and 
 [OIII] in two separate spectral windows. 
 These were taken in Mexico with the 2.1~m San Pedro Martir Telescope and Manchester 
 Echelle  spectrograph in November 2014. This was supplemented by high dispersion 
 spectra of HD~185806 on the 1.2~m Mercator telescope on La Palma, Canary
 Islands in July 2021 using the HERMES instrument, confirming it as an M~4 late
 type star.
 
 \begin{figure*}
	\includegraphics[width=17cm]{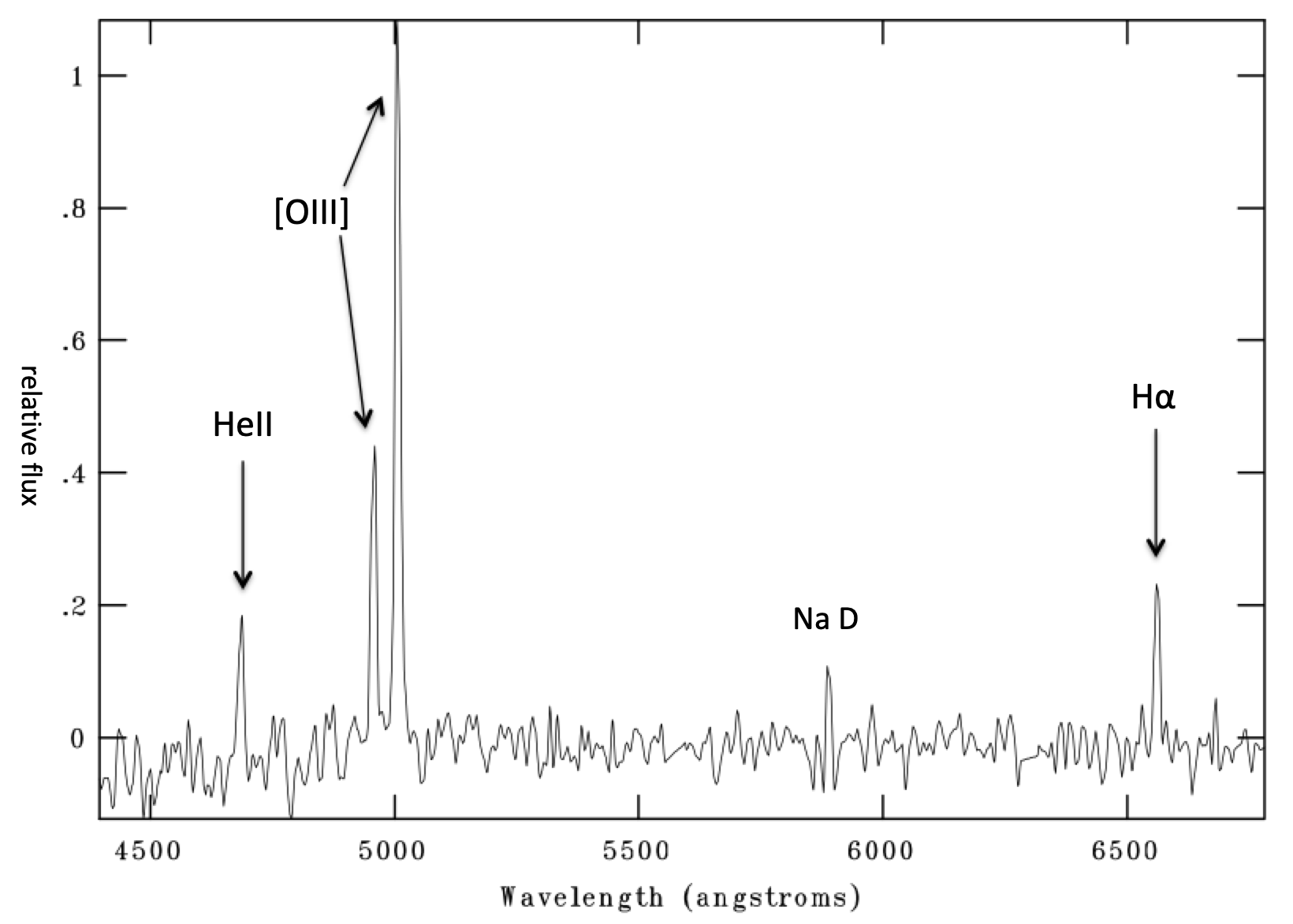}
    \caption{Low resolution 1-D optical spectra of WHTZ~1 taken with the 
    dedicated 2SPOT System in Chile. The [OIII] and 
    high excitation He~II 4686\AA~ emission lines are seen in 
    the blue, indicating it is a high excitation PN. 
    Weaker H$\alpha$ is evident in the red. The not fully cancelled bright 
    Sodium-D night sky emission line is also indicated.}
    \label{fig:2spot}
\end{figure*}

We present in Fig.~\ref{fig:2spot} new, low-resolution, optical spectra of
the nebula taken with the 2SPOT\footnote{for details on the 2SPOT amateur 
spectroscopy consortium see: \url{https://2spot.org/EN/}} Ritchey-Chrétien RC12 12-inch 
telescope in Chile equipped with the very stable Alpy 600 spectrograph from Shelyak Instruments 
(resolution 600).  The final 1-D spectrum was obtained after combining 
5$\times$1200~second exposures from the 12$^{th}$ and 10$\times$1200~second exposures from 
15$^{th}$ September 2022. Not only are the 
[OIII] optical emission lines visible but clearly so is the high excitation He~II
line at 4686\AA~. In the red the H$\alpha$ can be seen but is not strong compared to 
[OIII].  The residual Sodium D (Na~D) line is from the night sky. The 2-D spectral images 
were reduced following the standard pipeline describe in Le D\^u et al. (in press). The 
presence of He~II indicates a high excitation nebula. This would be an unprecedented 
detection for any optical bow-shock, e.g. see 
\citet{1999ApJ...526..274H} for attempts at such detection.

\subsection{Proper motion arguments}

Gaia \citep{2016A&A...595A...1G} DR3 data have now provided reliable proper 
motion and parallax data for both HD~185806 and the CSPN. This provides the 
opportunity to examine the proper motion vectors of these stars with respect to the 
nebula. As has been shown by \citet{2012A&A...538A.108P} with a MIR selected sample 
of bow-shock candidates, the earlier Hipparchus proper motion vectors of the stars 
are well aligned with their bow-shocks, i.e. they generally point in
the direction of the bow-shock. For HD~185806 no Gaia proper motion vector data were 
provided by \citet{2022MNRAS.515.1544S}, nor overlaid on their figures. 
The proper motion vector is added in Fig~\ref{fig:WISE} which points almost directly 
South and is not aligned with the major symmetry axis
of the nebula but is offset by $\sim$50~degrees.

\section{Broader Geometric considerations}

Visual inspection of the available, deep, narrow band imagery 
indicates the proposed blue CSPN is the actual 
ionising source for the nebula. It is located within 5~arcseconds of the 
intersection of the major and minor axis best fit by eye to the nebula's 
overall oval shape as would be expected for 
any true CSPN. This is also the case for the WISE MIR imagery where the 
optical enhancements on the NE and SW sides are reflected in a two 
component but uneven MIR intensity arc structure and hollow interior in 
the 22$\mu$m band-4 WISE data where the CSPN resides. The 9$\times$9~arcmin 
band-4 WISE image in Fig~\ref{fig:WISE} 
is dominated by HD~185806. This makes it difficult to assess the MIR emission 
component from the nebula as it is so close to the bright star, at least in a 
projected sense. Apart from the compact WISE nebula there is no other obvious 
MIR emission across the  9$\times$9~arcmin field shown. 
Contour fits \citep[as also done by][]{2022MNRAS.515.1544S} can 
give a false impression of the actual nebula emission close to such a 
bright star so examination of the MIR imagery itself is needed. There is 
a hint of W4 WISE emission beyond the star to the SE. We believe 
this shows the faint nebula extension expected here, assuming it 
reflects the faint outer envelope of the oval nebula seen to the NW in 
the optical and MIR imagery.

A further piece of contrary evidence to a bow-shock 
interpretation can be seen in Fig.~3 of \citet{2022MNRAS.515.1544S} where, at the 
extreme Northern corner of the [OIII] emission mosaic, there appears to be a faint 
emission arc parallel to the enhanced Northern edge of the main nebula but 
3~arcminutes away from the CSPN. This could be [OIII] emission associated with a possible
AGB halo of the PN given its geometry and alignment. In Fig~\ref{fig:Deep-Halpha} we show that indeed there is a faint outer 10~arcmin dimension halo visible, in this case in H$\alpha$.
 
\subsection{The bow-shock interpretation}
Bow-shocks can form both around hot, young stars where strong stellar 
winds slam into the surrounding interstellar medium or more commonly 
where runaway typically O/B stars are ploughing into a dense ISM.
Examination of the best extant image examples of bow-shocks around 
mostly runaway stars have little in common with the observed nebular 
features seen here. See the compilations,  discoveries and imagery 
reported by \citet{2012A&A...538A.108P} and later by 
\citet{2016ApJS..227...18K} and 
\citet{2017AJ....154..201K}. These
show  major, new samples of Galactic candidate stellar bow shocks found 
on Spitzer and WISE MIR data for runaway stars. All 244 examples listed 
by \citet{2012A&A...538A.108P}  
were from O and B type stars and no M-type stars, like 
HD~185806, were found in their entire sample. However, 
\citet{2012A&A...537A..35C} 
provide a catalogue of “wind-ISM” bow shocks and “wind-wind” 
interactions detected around AGB stars but only as seen in the far 
infrared with Herschel/PACSs imagery \citep{2001ESASP.460...13P} and not
at these shorter wavelengths.
 
Most bow-shocks exhibit very open angles on the forward curve and 
are well separated and in front of the associated star.  The only example
found that even remotely resembles the case here is associated with the 
binary star BZ~Cam but the prominent striations seen are lacking in that 
case and there is a prominent, if faint, more forward bow-shock too. 
 
The 3 main arguments to support a bow-shock interpretation made by 
\citet{2022MNRAS.515.1544S} while
seemingly plausible, lack firm observational evidence and are more 
descriptive than numerical. The claim of a bow-shock structure being 
"clearly depicted" in the WISE imagery is based on contour fits to the WISE
w4 22$\mu$m data that are affected by the intensity and close 
proximity of HD~185806 in the MIR. Close examination of the actual MIR 
imagery does not bear this out while they themselves admit to no 
evidence of a bow shock in the optical band as might be expected. 
They multiply claim that HD~185806 is "surrounded" by the nebula
when there is clearly almost no projected overlap between the star and the main 
body of the nebula region.

The star HD~185806 is not aligned with the major axis of the fitted ellipse that
best describes the nebula but is $\sim$15~degrees off, while the star itself is also 
located $\sim$1.5~arcmin to the SE of the nebula centre. The expected faint SE 
extension of the nebula to match that seen to the NW is largely obscured by the 
brightness of HD~185806  but can just be discerned at the edges. The Gaia proper motion 
vector is poorly aligned with the major axis of the nebula.

\section{Distance estimates and Kinematic age for the nebula}

PN WHTZ~1 was included in the calculated Surface Brightness 
radius (SB-r) relation PN distances paper of
\citet{2016MNRAS.455.1459F} as 
entry 307. The published distance from this robust PN statistical 
distance indicator is 2.49$\pm$0.71~kpc. This compares with a Gaia DR3 
parallax based distance for the assumed CSPN of 1.83~kpc \citep{Bailer-Jones2021}. This just agrees to within the admittedly large 20-30\% errors typically reported for the 
SB-r technique so the PN and CSPN distances are compatible. The robust 
Gaia distance for HD~185806 is 910~pc \citep{Bailer-Jones2021}.

The expansion velocity of the gas inferred from the data reported by 
\citet{2022MNRAS.515.1544S} is 
taken as $\sim57\pm10$~km $\mathrm{s^{-1}}$ based on an average of the positive and negative
velocity components reported at multiple points across the nebula and 
from the H$\alpha$, [NII] and [OIII] emission lines sampled. This is in 
the accepted range for a PN. Assuming this has been effectively constant 
since the envelope ejection from the progenitor star that created the PN and taking the 
physical size of the PN to be 1.7$\times$1.2~pc, as estimated by taking the PN angular size in radians and the Gaia DR3 distance for the assumed CSPN of 1.8~kpc, yields a 
kinematic age of the nebula of $\sim$25,000 years (see Tab.~\ref{tab:coord}). This is within
the range of PN kinematic ages, if at the higher end, but certainly not 
without precedent for such a faint and evolved PN \citep{2022ApJ...935L..35F}.

\section{Dust and extinction in the local environment}

The WISE data shows there is no further MIR emission in a large 
9$\times$9~arcmin area around WHTZ~1 - it is an isolated 
3$\times$2~arcmin MIR emission region. Even on one degree scales there
is very little MIR emission at this Galactic location and what there is 
is extremely faint and very diffuse. Likewise, the Schelgel, Finkbeiner and 
Davis dust maps \citep{1998ApJ...500..525S} show little dust in this 
zone while the large-scale SHASSA \citep{2001PASP..113.1326G}
H$\alpha$ gaseous emission maps show no significant ionised regions in 
the vicinity. This is not surprising given the nebula is 10~degrees 
latitude above the Galactic mid-plane. The ISM is not significant here 
and the extinction is assumed modest. 

The 3D dust map by \citet{Green2019} gives a low E(B-V) value of 0.33, quantitively 
confirming the impressions of a relatively low extinction environment. This
explains why the [OIII] narrow-band images and spectroscopic emission 
lines are strong as these data are not significantly extincted. 
In low extinction environments the [OIII] lines can be much stronger than 
H$\alpha$ emission in PNe. This is seen in many higher Galactic latitude 
examples such as Abell~78 and as is the case for this emission object.

\section{Corollaries with other known PNe}

The morphological structure of the nebula is typical of many 
elliptical shaped PNe. There are a many known PN corollaries, including a
recently discovered PN found by the French Amateur group, StDr~47 (HASH 
ID~32596), LTNF~1 (HASH ID~677) reported by 
\citet{1995ApJ...441..424L}, Kn~121 (HASH ID~23344) and Abell~36 (HASH ID~988) where all their positions, sizes, imagery etc can be located. 
Deep, narrow band amateur imagery of all of these PNe corollaries can also be found here: 
\url{https://www.imagingdeepspace.com/peter-goodhew-planetary-nebulae-images.html}. 

\begin{figure}
	\includegraphics[width=\linewidth]{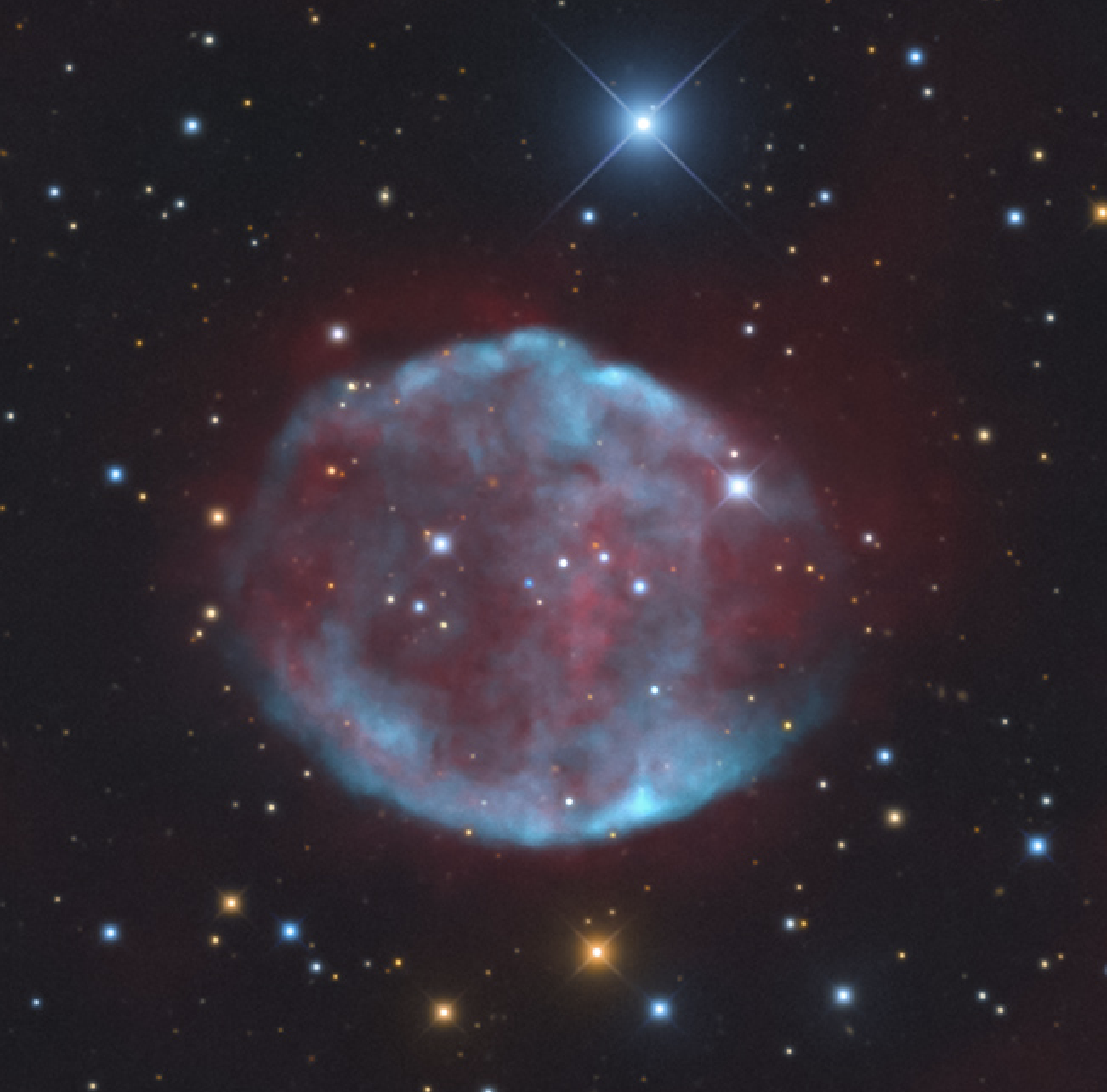}
	\includegraphics[width=\linewidth]{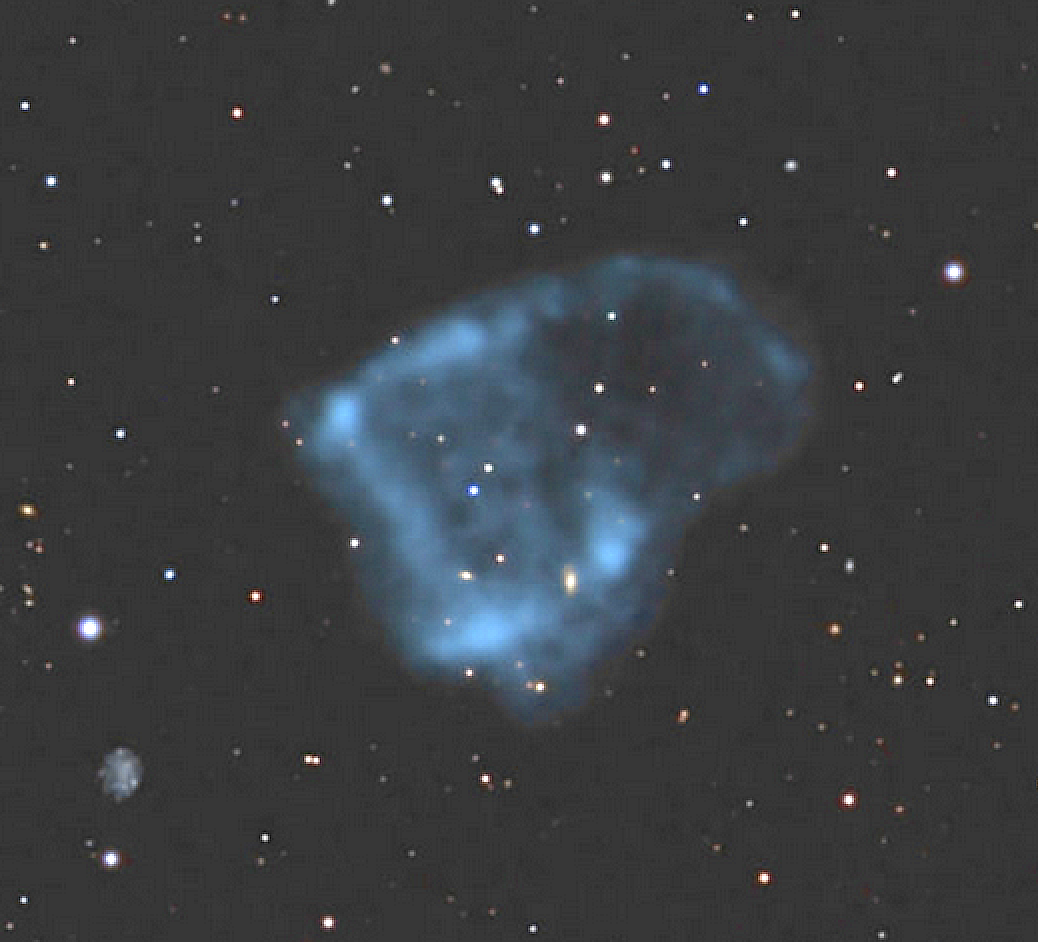}
    \caption{Deep, narrow-band H$\alpha$ and [OIII] amateur images of PN 
StDr~47 (HASH ID~32596, top  with PN major axis 270~arcsec) and LTNF~1 (HASH ID~677, bottom - 
    with PN major axis 230~arcsec) as examples of excellent morphological
    PNe corollaries to WHTZ~1. This includes the presence of faint blue 
    CSPNe at their geometric centres and even, in the case of StDr~47, a 
    bright, unrelated star nearby.}
    \label{fig:StDr47}
\end{figure}

These PNe share many of the features of WHTZ~1. They have an elliptical 
morphology, enhanced opposing edges, similar internal structure, a faint 
blue CSPN at the geometric centre and even, in the cases of StDr~47 and Kn~121, unrelated 
nearby bright stars. Two examples are shown for comparison purposes in 
Fig~\ref{fig:StDr47} and are extracted from deep narrow-band H$\alpha$ 
and [OIII] amateur images\footnote{see 
\url{http://planetarynebulae.net/EN/page\_np.php?id=845} for StDr~47 and 
\url{https://www.imagingdeepspace.com/ltnf1.html} for LTNF~1)}. There 
are many examples of PNe being in close projected proximity to unrelated 
bright stars as might be expected given the majority of PNe are located 
in the Galactic plane. Perhaps the best known example is Abell~12 the so-called "hidden 
planetary", a small PN in the constellation of Orion heavily obscured by the glare of 
4$^{th}$ magnitude star Mu Orionis in the optical but nicely seen in the WISE MIR imagery.

\section{Summary of observed properties and assessment}

We provide below the main accumulated arguments against the bow-shock
interpretation and supporting the PN nature of the observed nebula emission.

\subsection{Arguments against the bow-shock scenario}

A brief summary of points concerning the bow-shock interpretation are as 
follows:
\begin{itemize}
    \item The morphology of the nebula is unlike other extant 
    examples in the literature of bow-shock nebulae in the optical
    \item The host star is not an O or B type star as found for most 
    other examples in the optical and MIR as reported by 
    \citet{2012A&A...538A.108P} 
    but is an M~4 late type star. This would be very unusual at these 
    wavelengths, although some are found around AGB stars in the far infrared with Herschel 
    \item The presence of an faint outer nebular halo fits a PN AGB halo interpretation but not a bow-shock
    \item The star's location is not aligned with the major axis of the elliptical nebula with a $\sim$15~degree offset
    \item The strength of [OIII] relative to H$\alpha$ can be explained by shocks but can also easily be explained in a PN interpretation and by the Galactic location out of the mid-plane so the blue emission lines are not heavily extincted
    \item The low extinction and lack of gas and dust in the general area means the ISM does not provide a dense medium for bow shocks to develop easily
    \item The stellar and nebula systemic velocities do not agree
    \item The Gaia proper motion vector of the star is not aligned with the supposedly trailing nebula (see Fig.~\ref{fig:WISE}) 
    but is off by $\sim$50~degrees
\end{itemize}

\subsection{Factors supporting a PN identification}

A brief summary of points supporting a PN classification is as follows:
\begin{itemize}
\item The morphology is PN like with many corollaries with known PNe (see Fig.~\ref{fig:StDr47})
\item There is a plausible blue, ionising star at almost the exact geometric centre of the nebula. This is a compelling argument for a PN origin for the nebula emission 
\item the optical spectrum is also compelling given the spectrum has a clear detection of He~II 4686\AA~ emission which requires a CSPN with a T$_{eff}$ of at least 50,000~K, implying a 
high excitation PN identification
\item The SB-r distance to the nebula and the CSPN parallax distance from Gaia are compatible within the errors
\item The expansion velocity, kinematic age and physical size are all within accepted ranges for PNe
\item The presence of a faint 10~arcmin outer halo - shown in our deep H$\alpha$ image (Fig~\ref{fig:Deep-Halpha}), is a feature also seen in many PNe
\end{itemize}

Based on all these combined observed characteristics and also applying the decision tree for PNe to the key observations provided in Fig.11 of \citet{Parker2022} leads us to a straightforward identification of WHTZ~1 as a high excitation PN.

In Tab.~\ref{tab:coord} we present the combined summary of all the observed and estimated parameters for the PN, proposed CSPN and for HD~185806.

\renewcommand{\arraystretch}{1.5} 
\begin{table*}
	\centering
	\caption{Summary Table of all observed and determined PN, CSPN and star HD~185806 parameters from this work and from the literature.}
	\label{tab:coord}
	\begin{tabular}{lccc}
	\hline
         Parameter & PN & CSPN & HD~185806\\ 
         \hline
         RA (J2000) & 19:40:43.84 & 19:40:$43.840^\bigtriangledown$ & 19:40:$47.56^\bigtriangledown$ \\
         DEC (J2000) & 02:30:31.80 & 02:30:$31.928^\bigtriangledown$ & 02:29:$28.64^\bigtriangledown$ \\
         Galactic longitude & 40.875 & 40.87508  & 40.86675\\
         Galactic latitude & -9.7764 & -9.77643 & -9.79839 \\
         Gaia stellar ID: 428968+ &  - & 4146940013184 & 3322306283904\\
         Gaia G-band (mag) & - & 16.887  &  7.606\\
         $B_j$ (mag)$^{\diamond}$  & -  & $\sim$16.94 &  $\sim$10.37\\
         R-band (mag)$^{\diamond}$ & - &  17.42 & 8.00\\
         $E (B - V)^\bigtriangleup$ & 0.33 & - & -\\
         $\mu_{\delta}$ $(\mathrm{mas yr^{-1}})^\bigtriangledown$ & - & $-7.78 \pm 0.06$ & $-10.28 \pm 0.03$\\
         $\mu_{\alpha}\mathrm{cos}\delta$ $(\mathrm{mas yr^{-1}})^\bigtriangledown$ & - & $-5.03 \pm 0.08$ & $-0.36 \pm 0.04$\\
         Distance (kpc) & $2.49\pm0.71^*$ & $1.83^{+0.27^\bullet}_{-0.21}$ & $0.91^{+0.04^\bullet}_{-0.02}$\\
         Velocity (km $\mathrm{s}^{-1}$) &  $-34\pm13$ (systemic)$^{\#}$ & - & $13.76\pm0.37$ (heliocentric)$^\bigtriangledown$\\
          & $57\pm10$ (expansion)$^{\#}$ & - & -\\
         Physical size major axis (pc) & $1.72^{+0.24}_{-0.21}$ & - & -\\
         Physical size minor axis (pc) & $1.19^{+0.17}_{-0.12}$ & - & -\\
         Nebula outer H$\alpha$ halo (arcmin) & $\sim$10 & - & -\\
         Kinematic age (yr) & $25000^{+15000}_{-10000}$ & - & -\\
         \hline
         \multicolumn{4}{l}{\small $^{\bigtriangledown}$from Gaia DR3 \citep{GaiaDR3_2022}} \\
         \multicolumn{4}{l}{\small $^{*}$from SB-r relation of \citet{2016MNRAS.455.1459F}} \\
         \multicolumn{4}{l}{\small $^{\#}$ estimated from data in \citet{2022MNRAS.515.1544S}}\\
         \multicolumn{4}{l}{\small $^{\diamond}$ taken from SuperCOSMOS photometry, see \citet{2001MNRAS.326.1279H}}\\
         \multicolumn{4}{l}{\small $^\bigtriangleup$ taken from \citet{Green2019}}\\
         \multicolumn{4}{l}{\small $^\bullet$ taken from \citet{Bailer-Jones2021}}
	\end{tabular}
\end{table*}

\section{Conclusions}
We have shown that the recently reported cocoon "bow-shock" nebula, 
supposedly trailing the bright, 9.4$^{th}$ magnitude star HD~185806 is 
actually a previously identified, faint, evolved, high excitation PN 
WHTZ~1 (Ra~7). This is confirmed by new, low resolution optical spectroscopy and 
multiple other pieces of evidence. The blue, ionizing CSPN, detected in Gaia 
DR3 \citep{GaiaDR3_2022} as star 4289684146940013184, has a reported parallax of 0.5548~mas 
from which \citet{Bailer-Jones2021} infer a distance of $\sim$1.8~kpc. This gives the PN a physical 
extent across the major axis of $\sim$1.7~pc, well within the range of known PNe. The 
expansion velocity and kinematic age are also compatible with a PN origin. The elliptical
morphology is also typical of many 
evolved PNe with LTNF~1, KN~121, Abell~36 and the recent discovery of PN StDr~47 being 
particularly apt corollaries. They all have faint blue CSPNe with very similar 
morphological structures and, in the case of StDr~47 and KN~1212, bright, unrelated 
close stellar projected neighbours. The bright star HD~185806 is 
merely a chance line-of-site projected neighbour situated in the foreground at 910~pc.

\section*{Acknowledgements}
We would like to thank the anonymous referee who helped us improve the paper.
QAP thanks the Hong Kong Research Grants Council for GRF research support
under grants 17326116 and 17300417. AR thanks HKU for the provision of a 
postdoctoral fellowship. This work has made use of data from the European
Space Agency (ESA) mission
{\it Gaia~} (\url{https://www.cosmos.esa.int/gaia}), processed by the {\it Gaia}
Data Processing and Analysis Consortium (DPAC,
\url{https://www.cosmos.esa.int/web/gaia/dpac/consortium}. Funding for 
the DPAC has been provided by national institutions, in particular the 
institutions participating in the {\it Gaia} Multilateral Agreement. We also thank 
Johannes Schedler of the Chart32 
team for the excellent Chart32 processed images used in this paper.

\section*{Data Availability}
The data underlying this article are available in the article itself and 
in its associated online material freely accessible from the HASH 
database found here: \url{http://hashpn.space} by simply entering the 
unique HASH ID number for each source as provided.



\bibliographystyle{mnras}
\bibliography{references} 





\bsp	
\label{lastpage}
\end{document}